\documentclass[11pt,a4paper]{article}
\usepackage{mathrsfs}
\usepackage{amsmath}
\usepackage{amssymb}
\usepackage{mathdesign}
\usepackage{dsfont}
\usepackage{esint}
\usepackage{youngtab}

\usepackage{amsmath,amssymb,amsfonts,a4wide,graphicx,bm,times,psfrag,wrapfig,sidecap}
\usepackage{cite}
\usepackage[colorlinks=true,linkcolor=black, citecolor=black,
urlcolor=black]{hyperref}
\numberwithin{equation}{section}
\makeatletter \let\old@startsection=\@startsection
\renewcommand{\@startsection}[6]
{\old@startsection{#1}{#2}{#3}{#4}{#5}{#6\mathversion{bold}}}
\makeatother

\def\Res{ \text{Res}}

\def\be{\begin{eqnarray}  }
    \def\ee{\end{eqnarray}}

\def\({\left(} \def\){\right)}
\def\<{\langle}
\def\>{\rangle}
\def\[{\left[}
 \def\]{\right]}
\def\tr{{\rm   tr} }

\newcommand\encadremath[1]{\vbox{\hrule\hbox{\vrule\kern8pt
\vbox{\kern8pt \hbox{$\displaystyle #1$}\kern8pt}
\kern8pt\vrule}\hrule}} \def\enca#1{\vbox{\hrule\hbox{
\vrule\kern8pt\vbox{\kern8pt \hbox{$\displaystyle #1$} \kern8pt}
\kern8pt\vrule}\hrule}}

  \usepackage{bm}

\def\ee{\end{eqnarray}}  

\def\({\left(} \def\){\right)}   \def\[{\left[} \def\]{\right]} \def\tr{{\rm tr} }

\def\Re{{\rm Re}}
\def\Im{{\rm Im}}

\begin{document}

\thispagestyle{empty}

\begin{flushright}

\end{flushright}

\vspace{1cm}
\setcounter{footnote}{0}

\begin{center}

\begin{center}

{\Large\bf Long Time Behvaior of the Kondo Model After a Quench}

\vspace{7mm}
Eldad Bettelheim  \\[5mm]

{\it Racah Inst.  of Physics, \\Edmund J. Safra Campus,
Hebrew University of Jerusalem,\\ Jerusalem, Israel 91904 \\[5mm]}

\end{center}

\abstract
We find the statistical weight of excitations at long times following a quench in the Kondo problem. The weights  computed  are directly related to the overlap between initial and final states that are, respectively, states close to the Kondo ground state and states close to the normal metal ground state. The overlap is computed making use of the Slavnov approach, whereby a functional representation method  is adopted, in order to obtain de        finite expressions.  
\end{center}
\section{Introcution}
The Bethe ansatz method, when applicable, typically allows for a full description of the equilibrium thermodynamics of certain strongly interacting quantum systems. The extension of the method to non-equilibrium, is rather an open problem. In this paper we shall apply the Bethe ansatz approach to the  Kondo problem, an epitomic example of a quantum integrable system. We shall treat that system out of equilibrium, but close to the ground state. Such a setting is achieved by considering, for example, the long time Loschmidt echo following a quench from the normal metal ground state to the Kondo state. 

The Loschmidt echo can be written as a the following object:\
\begin{align} 
L(\gamma) =\sum_k e^{-\gamma } |\<i|E_k\>|^2,
\end{align}
where $\gamma$ is imaginary time.

In the Bethe ansatz approach, the states $|E_k\>$ are enumerated by solutions of the Bethe ansatz equations \cite{Wiegmann:First:Kondo,AndreiL:First:Kondo,Wiegmann:Tsvelick:Review:Kondo,Andrei:Lowenstein:Review}. The Bethe ansatz equations are a nonlinear set of equations for  a set of complex numbers, called rapidities. These describe the spins of the electrons. Indeed, in the Kondo approach charge and spin may be treated separately. The charge sector is equivalent to  a set of spinless fermions, while the spin sector is described by the Bethe ansatz of a non-homogeneous Heisenberg system.  In the thermodynamic limit, the complex rapidities of the spin sector  form line densities on the lines $\frac{\imath}{2}n \mathds{R} ,$ where $n\in\mathds{Z}$.  The system of densities on these lines is denoted by $\bm \sigma$ . In addition, one requires  a set of integers ${\bm n}$ to describe the eigenstate of the charge system. These numbers are related to the spinless fermion charge sector momenta . We shall, from now on, then write   $|\bm \sigma,\bm n\>$ instead of $|E_k\>,$ for an eigenstate of the full Kondo problem. For more details on this procedure one may consult Refs.  \cite{Wiegmann:First:Kondo,AndreiL:First:Kondo,Wiegmann:Tsvelick:Review:Kondo,Andrei:Lowenstein:Review}  . In addition,  Ref. \cite{Bettelheim:Kondo} provides details of the current non-equilibrium formal approach using the same notations.  

One can then pass from a sum over eigenstates to an integral over all possible line densities of the rapidities, $\bm\sigma$. The price is, as to be expected, an entropy term. The result is that the Loschmidt echo, for example, is expressed as:
\begin{align}
L(\gamma)= \int \exp\left[S-\gamma E+\mathcal{A}  \right] \mathcal{D}{\bm \sigma}\mathcal{D}{\bm n},\label{QuenchedActionEqIntegral}
\end{align}
where the entropy $S$, and the energy $E$\ are functionals of the rapidity density, $\bm\ \sigma$. The logarithm of the modulus square of the  overlap between the initial state and the eigenstate $|E_k\>$ becomes also a functional of $\bm \sigma$, which is denoted by  $\mathcal{A}$:
\begin{align}
\mathcal{A} = 2\log |\<i|\bm{\sigma}, \bm n\>|. 
\end{align}
The quantities $S$ and $E$ are familiar from equilibrium and are thus well-known functionals of $\bm{\sigma},  $  due to their appearance already in equilibrium thermodynamics, and their expression in this context are given in  Refs.  \cite{Wiegmann:First:Kondo,AndreiL:First:Kondo,Wiegmann:Tsvelick:Review:Kondo,Andrei:Lowenstein:Review} , while a discussion of the computation of $\mathcal{A}$ in general in the Kondo problem was a subject of Ref. \cite{Bettelheim:Kondo}. Here we shall concentrate on the case where $i$ is close to the ground state of the normal metal, while ${\bm \sigma}$ is close to the ground state of the Kondo problem 

Since the number of particles is large, the integral in (\ref{QuenchedActionEqIntegral}) is dominated by the saddle point. The saddle point equations read 
\begin{align}
\frac{\delta F}{\delta \bm \sigma} = \gamma^{-1}\frac{\delta \mathcal{A}}{\delta \bm \sigma} \label{QActionEq},
\end{align}
 where $F=E-\gamma^{-1} S.$  The relation (\ref{QActionEq}) was first noted  in \cite{Caux:Essler:Time:Evolution:After:Quench,Caux:Lieb:Liniger:Quench,Caux:Heisenberg:Ising:Quench}, and was termed the quench action approach. The focus of this paper is to
find the right hand side of this equation in the problem described above. 
\section{Bethe Ansatz Description of Low-lying States}
As mentioned in the introduction, we shall use the Bethe ansatz description of the initial and final states in a quench. To do so, we shall need to recount some facts about the Bethe-ansatz diagonalization of the Kondo problem  \cite{Wiegmann:First:Kondo,AndreiL:First:Kondo,Wiegmann:Tsvelick:Review:Kondo,Andrei:Lowenstein:Review} . 

First we note that an eigenstate of the Kondo problem can be obtained by first writing an eigenstate for the spin system associated with the electron. This state is written as $|\bm \sigma\>$, where $\bm \sigma$ is the density distribution of rapidities that determines the spin state. Once the spin state has been determined the charge sector is described by spinless free fermions, which undergo a phase shift at the impurity, the magnitude of which depends on the spin system, such that we write $\delta (\bm \sigma)$ for this phase shift. The expression for the phase shift may be found in   \cite{Wiegmann:First:Kondo,AndreiL:First:Kondo,Wiegmann:Tsvelick:Review:Kondo,Andrei:Lowenstein:Review,Bettelheim:Kondo}. To stress this fact we write $|\bm n,\delta(\bm\sigma)\>$ for the charge system, where $\bm n$ denotes a set of integers $\bm n=\{n_i\}_{i=1}^N$, which are related to the wave vectors of the spinless electrons through $k_i=\frac{2\pi n_i}{L}$. Spin charge separation is embodied in the relation:  
\begin{align}
|\bm \sigma, \bm n\> = |\bm \sigma\>\otimes |\bm {n}, \delta (\bm \sigma)\>.
\end{align}

The spin-state $|\bm \sigma\>,$ is a highest state in a spin multiplet. Two states must be in the same multiplet and have the same spin projection to have an overlap. The rapidities in $\bm \sigma$ arrange themsleves in strings. We shall be interested only in the case where the strings are of length $1$, such that all rapidities in $\bm \sigma$\ are actually arranged on the real axis. The distribution of rapidities may have holes. If the initial state is of that form then the final state is also of that form. This is the case, for example, when one applies a small magnetic field. The ground state with a magnetic field is a state where holes are introduced into the ground state.  If the number of electron is $N$\ and there are $M$\ rapidities then the spin of the state is $N-M$.

We denote by $\bm v$ the set of rapidities $v_i$ of the initial, normal, and $\bm u$ the set of rapities, $u_i$, of the final Kondo state
\begin{align}
\bm v = \{v_i\}_{i=1}^{M+1},\quad  \bm{u}=\{u_i\}_{i=1}^{M+1}.
\end{align} 
The free fermion problem is described in the Kondo approach by setting \begin{align}v_{M+1}=-g^{-1}-\frac{\imath}{2}+\delta\end{align} and letting $\delta $ tend to $0$ at the end, while the rest of the rapidities $\{v_i\}_{i=1}^M$, obey the non-interacting Bethe ansatz equations. In fact, we shall denote by $\sigma^{(0)}$ the distribution of the normal state rapidities for $M=\frac{N}{2}$, namely for a singlet state (from now on $N,$  the number of electrons, is assumed to be even for convenience). The density,  $\sigma^{(0)},$  is given by:
\begin{align}
\sigma^{(0)}(x)=\frac{N}{2\cosh\pi x},\label{FFSigma}
\end{align}
 We also define the set of rapidities $\bm v^{(0)}$ as the set of rapidities corresponding to a density given by (\ref{FFSigma})

The addition of the impurity causes an order $1$ change in the density describing the distribution of the rapidities, as well as the addition of any order $1$ number of holes in the distribution of rapidities, such that all distributions of rapidities on the real axis in question in this paper are given by $\sigma^{(0)}$ of Eq. (\ref{FFSigma}), to leading  order (namely to order $N$), and all deviations from this distribution are of order $1$. 

We shall also be interest in the function $f_{\delta \sigma}$ which denotes the shift in the location of the final state rapidities due to the variation of the density:
\begin{align}
\delta v_i = f_{\delta\sigma}(v_i).
\end{align}
Varying the density is required to compute the variation derivative on the RHS\ of (\ref{QActionEq}). The shift of the location of the rapidities, $f_{\delta \sigma}$, under such a variation of the rapidity distribution is related to the change in the line density of the rapidities in the following fashion:
\begin{align}
\delta \sigma^{(\bm u)}(x) =-\partial_x( \sigma^{(\bm u)}(x) f_{\delta \sigma}(x)).\label{fdeltasigmarelation}  
\end{align}
Since $\delta\sigma\sim 1$ and $\sigma \sim N$ then $f_{\delta \sigma}\sim 1/N$.

We shall also need the set (with repetitions) of inhomogeneities, $\bm z$, which is given by
\begin{align}
\bm z =\Bigl(\frac{\imath}{2} -\frac{1}{g},\overset{\times N}{\overbrace{\frac{\imath}{2},\dots,\frac{\imath}{2}}}\Bigr). 
\end{align}
We define four  sets  (with repetitions), $\bm c^{(i)},$ where $i$ can take  one of four values $i\in\{NK,KK,NN,0\}$: 
\begin{align}
{\bm c}^{(i)}=\lim_{\bm u' \to \bm u ,\bm v' \to \bm v ,\bm v'^{(0)}\to\bm v^{(0)}}\left\{\begin{array}{lr} \bm u\cup\bm v & i=NK \\ \bm u'\cup\bm u & i=KK\\  \bm v'\cup\bm v & i=NN \\  \bm v'^{(0)}\cup\bm v^{(0)} & i=0    \end{array}\right., 
\end{align}
For  any set with repetitions , $\bm S,$ we define the polynomial $Q_{\bm S,}$ as the monic polynomial with zeros on the set $\bm S$ and only on $\bm S$:  
\begin{align}
Q_{\bm S} (x)= \prod_{x\in \bm S} (z-x).
\end{align}

\section{Bethe Ansatz Description of the Overlaps and Norms}

The overlaps and norms are  given by a Slavnov determinant\cite{Slavnov} given here in a form due to Kostov and Matsuo in Ref.  \cite{Kostov:Inner:Product:Domain:Wall} : 
\begin{align}
\prod_j\left(-Q_{\bm z}(b_j+\imath)Q_{\bm z}(a_j)\right)^{-1}\det B^{(\bm a\cup \bm b )}=\<\bm a|\bm b\> \label{OverlapasDet}, 
\end{align}
where the matrix $B$ is given by:
\begin{align}
B^{(\bm c)}_{kj} =\delta_{i,j} -\frac{Q_{\bm z}(c_k)Q_{\bm c}(c_k+\imath)}{Q_{\bm c}(c_k)Q_{\bm z}(c^{(i)}_k+\imath)} \frac{1}{c_k-c_j+\imath}.\label{Bexpression}
\end{align}
This particular form of $B$ was given in Ref. \cite{Kostov:Bettelheim:SemiClassical:XXX}. We shall naturally denote:
\begin{align}
B^{(i)}\equiv B^{(\bm c^{(i)})}.\label{brevity}
\end{align}
Computing the inverse of $B^{(i)}$ allows to compute the variation of the logarithm of the determinant, which is the object we are interested in. The inverse is encoded in $\mathcal{R}^{(i)}$ as follows:
\begin{align}
\mathcal{R}^{(i)}(z,c^{(i)}_k)=\sum_k \frac{(B^{(i)-1})_{jk}}{z-c_k^{(i)}}.\label{Rdefinition}
\end{align}
It was established in \cite{Bettelheim:Kondo} that $\mathcal{R}$ so defined solves the following integral equation:
\begin{align}
&\mathcal{R}^{(i)}(z,w) - \oint e^{-N\varphi^{(i)}(z')}   \frac{\mathcal{R}^{(i)}(z'+\imath,w)dz'}{2\pi\imath(z'-z)} = \frac{1}{z-w}-\frac{ \gamma ^{(i)}(w)}{z+\frac{\imath}{2}+g^{-1}}.\label{Equation4R}
\end{align}  
where:
\begin{align}
e^{-N\varphi^{(i)}(z)}=\frac{Q_{\bm c^{(i)}}(z+\imath)}{Q_{\bm z}(z+\imath)}\frac{Q_{\bm z}(z)}{Q^{(i)}_{\bm c}(z)} \label{etothephiExp}
\end{align}
and the contour of integration surrounds here and below the real axis. Here the contour of integration also surrounds the point $w$.
(In Ref. \cite{Bettelheim:Kondo}, the contour of integration surrounded also the lines $c_{N+1}$, but we adopt a different convention here and make the requisite adjustments associated with this change.)

The function $\gamma^{(NK)}(w)$ is  to compensate for the fact that  the contour integral in (\ref{Equation4R}) does not surround  the rapidity $c_{N+1}=-g^{-1}-\frac{\imath}{2}+\delta.$ As such this function is given by: 
\begin{align}
\gamma^{(NK)}(w)=-\underset{z\to c_{N+1}}{\Res}e^{-N\varphi(z)}\mathcal{R}^{(NK)}(c_{N+1}+\imath,w).\label{gammaessenece}
\end{align}
For
$i=KK$ there is nothing to compensate for and the contour surrounds all rapidities, as such  $\gamma^{(KK)}=0$.  

A more useful property that allows to compute  $\gamma^{(NK)}(w)$ is that the following condition must be  satisfied (see Ref. \cite{Bettelheim:Kondo}):
\begin{align}
\mathcal{R}^{(NK)}(-g^{-1}+\frac{\imath}{2},w)=0\label{gammacondition}.
\end{align}
As we shall see below this condition allows to compute $\gamma^{(NK)}$ more readily than (\ref{gammaessenece}), since the former may be applied after $\delta \to0,$ while the limit procedure must be applied explicitly to make use of the latter.  

\section{The Integral Kernel}

We shall want to find a form for $e^{-N\varphi^{(0)}(z)}$ valid in the thermodynamic limit. To that aim, we first find the function whose jump discontinuity is (\ref{FFSigma}) up to a constant multiplicative factor. Namely we look for the following integral:
\begin{align}
\int_{-\infty}^\infty \frac{1}{(z-x')\cosh(\pi x')}.\label{SecIntegral}
\end{align}
The solution of this integral is facilitated by making use of the Hopf decomposition of the secant function: 
\begin{align}
\frac{  2\pi}{\cosh(\pi x)}=\psi ^{(0)}\left(\frac{\imath x}{2}+\frac{3}{4}\right)-\psi ^{(0)}\left(\frac{\imath
   x}{2}+\frac{1}{4}\right)+\psi ^{(0)}\left(-\frac{\imath
   x}{2}-\frac{3}{4}\right)-\psi ^{(0)}\left(-\frac{\imath
   x}{2}-\frac{1}{4}\right),\label{HopfDecomposeSech}\end{align}
where $\psi^{(0)}(x)=\frac{d}{dx}\log(\Gamma(x)).$ As a function of $z$, the integral in (\ref{SecIntegral}) is analytic in the entire space except at the real axis where it has a jump discontinuity of $\frac{-2 \pi \imath}{\cosh(\pi x)}$. In view of (\ref{HopfDecomposeSech}) such a  function is given by the right hand side of:
\begin{align}
\int_{-\infty}^\infty \frac{1}{(z-x')\cosh(\pi x')}=\imath \times\left\{\begin{array}{lr} \psi ^{(0)}\left(-\frac{\imath
   z}{2}+\frac{1}{4}\right)-\psi ^{(0)}\left(-\frac{\imath
   z}{2}+\frac{3}{4}\right) & \Im(z)>0 \\ \psi ^{(0)}\left(+\frac{\imath z}{2}+\frac{3}{4}\right)-\psi ^{(0)}\left(+\frac{\imath
   z}{2}+\frac{1}{4}\right) & \Im(z)<0 \end{array}\right. .
\end{align}  
Which gives
\begin{align}
\exp\left(\int^z\int_{-\infty}^\infty \frac{1}{(z-x')\cosh(\pi x')}\right)=C\times\left\{\begin{array}{lr} \frac{\Gamma^2\left(-\frac{\imath
   z}{2}+\frac{3}{4}\right)}{\Gamma^2(-\frac{\imath z}{2}+\frac{1}{4})} & \Im(z)>0 \\ -\frac{\Gamma^2\left(\frac{\imath
   z}{2}+\frac{3}{4}\right)}{\Gamma^2(\frac{\imath z}{2}+\frac{1}{4})} & \Im(z)<0 \end{array}\right. ,\label{ExpIntSec}
\end{align}
for some unimportant constant, $C$. 

We thus find:
\begin{align}
&\varphi^{(0)}(z)=\log\nonumber\frac{\left(z+\frac{\imath}{2}\right)}{\left(z-\frac{\imath}{2}\right)}-\\&-2\log\left(\begin{array}{lr} \frac{\Gamma\left(-\frac{\imath
   z}{2}+\frac{1}{4}\right)\Gamma(-\frac{\imath z}{2}+\frac{5}{4})}{\Gamma^{2}(-\frac{\imath z}{2}+\frac{3}{4})} & \Im(z)>0\\  \frac{\Gamma^{}\left(-\frac{\imath
   z}{2}+\frac{5}{4}\right)\Gamma^{}(\frac{\imath z}{2}+\frac{1}{4})}{\Gamma^{}(-\frac{\imath z}{2}+\frac{3}{4})\Gamma^{}\left(\frac{\imath
   z}{2}+\frac{3}{4}\right)} & 
-\imath<\Im(z)<0 \\  \frac{\Gamma^{2}\left(\frac{\imath
   z}{2}+\frac{1}{4}\right)}{\Gamma^{}(\frac{\imath z}{2}-\frac{1}{4})\Gamma^{}\left(\frac{\imath
   z}{2}+\frac{3}{4}\right)} & \Im(z)<-\imath \end{array}\right) .\label{varphi0explicitCases}
\end{align}
Note that $\varphi^{(0)}(x-\imath0^+)$ is real on the real axis.

\subsection{Fourier Transform of Kernel}

The function  $e^{-N\varphi^{(0)}(z)}$ (or rather this function up to $O(1)$\ corrections in the exponent) appears as the Kernel in the main integral equation Eq.  (\ref{Equation4R}) we need to solve. It turns out that working in Fourier space facilitates solving this equation. We thus define $\tilde G$ through the Fourier transform of the jump discontinuiy of    $e^{-N\varphi^{(0)}(z)}$, where the jump discontinuity is defined as the value of    $e^{-N\varphi^{(0)}(z)}$ just below the real axis minus the value just above, respectively. Explicitly,  we define $e^{-N\tilde{G}^{(0)}(P)}  $ as:
\begin{align}
e^{-N\tilde{G}^{(0)}(P)} = \oint e^{-N(\varphi^{(0)}(z) + \imath z P)}\frac{dz}{2\pi},\label{Gdef}
\end{align}
where the integral is to be taken counterclockwise around the real axis. The variable $P$\ is related to the usual Fourier space variable, $p$, by:
\begin{align}
P=\frac{p}{N} .
\end{align}

When referring to the contribution of the Fourier integral (\ref{Gdef}) only above or below the real axis to $e^{-N\tilde{G}^{(0)}}$ we shall make use of $e^{-N\tilde{G}_\pm^{(0)}}$, respectively:
\begin{align}
e^{-N\tilde{G}_\pm^{(0)}(P)}=\mp\int_{-\infty\pm\imath0^+}^{\infty\pm\imath0^+} e^{-N(\varphi^{(0)}(z) + \imath z P)}\frac{dz}{2\pi}.
\end{align}  
We shall also denote as $\varphi_\pm^{(i)}(z)$ as the analytical continuation of $\varphi^{(i)}(z)  $  from above or below the real axis, respectively. 

After defining the Fourier transform we now proceed to study its behavior.

The Fourier integral may be performed by the saddle point method. We first take care of the part of the contour integral above the real axis. For large  $P$ we find three saddle points, two of which are related to the expansion of $\varphi_+^{(0)}$ around the points $\pm \frac{\imath}{2}$. These saddle points are:
\begin{align}
z^{(A)}_* =\imath\left(\frac{1}{2}-\frac{1}{ P}+\dots\right),z^{(B)}_* =-\left(\frac{\imath}{2}+\frac{1}{ P}+\dots\right) \quad
\end{align}
A third saddle point, $z_*^{(C)}$, has a larger negative imaginary value, and is irrelevant to the calculation since the steepest descent contour does not pass through it.  

 It  can be ascertained that the steepest descent contour passes through $z_*^{(B)}$, which we denote simply by $z^+_*$. This produces a large and positive $\tilde {G}_+,$  for large $P$:
\begin{align}
\tilde{G}_+^{(0)}(P)=\varphi^{(0)}(z_*)+\imath z^+_* P+\dots=P+\log\left(-\frac{\pi^2}{e4P^3}\right)+\dots
\end{align}

For $P$ negative it is possible to deform the contour to infinity in the upper half plane without obstruction, where the integrand of the Fourier integral tends rapidly to zero, so the $\tilde{G}_+^{(0)}(P) $ may be considered positive and infinite for negative $P$. 
In fact,  $\tilde{G}_+^{(0)}(P) $ , diverges as $P$ tends to zero from above. For small positive $P$, the  saddle point turns out to be at large $z$. Indeed, for 
\begin{align}
z_*^{(j)}=\left(\frac{\imath}{ P}\right)^{\frac{1}{3}}=\frac{{\rm sign}(P)}{|P|^{1/3}}e^{\imath \left(\frac{\pi}{6}+(j-1)\frac{2\pi}{3}\right)},
\end{align}
the saddle point condition $\partial_z \varphi^{(0)}=-\imath P$ is satisfied approximately, and $z$ is large. Here $j$ runs from $1$ to $3$ . The value of the action at these saddle points is given by:
\begin{align}
\varphi_+^{(i)}(z_*^{(j)}(P))+\imath P z_*^{(j)} (P)= \frac{3|P|^{2/3}}{2}e^{\imath \frac{2\pi}{3} j}.
\end{align}The steepest descent contour passes through $z_*^{(0)}$, where action diverges to positive infinity as $P$\ tends to zero from above.

{ 

As $P$ decreases from infinity, the saddle points $z_*^{(A,B)}$ continue moving on the real axis until some critical  $P$ at which point a bifurcation occurs. Before this point, at the  value $P_*^+=2\pi$, where  $z_*^{(B)}(P_*)=0 $ (one can ascertain from (\ref{varphi0explicitCases}) that indeed $\varphi'^{(0)}_+(0)=-2\pi\imath$) the function $\tilde{G}$ obtains a minimum, as can be seen by expanding $\tilde G(P_*(z))=\varphi^{(0)}(z)+\imath z P(z)$  $z$ around $z=0$. One can check that this is a global minimum by further studying the function $\tilde{G}(P)$. The expansion of  the  Fourier action , $\varphi^{(0)}(z)+\imath zP$, around $P=2\pi $ and $z=0$ is as follows:
\begin{align}
\varphi_+^{(0)}(z)+\imath zP=-d_0+\imath (P-2\pi)z+4(2\mathcal{G}-1)z^2+\dots,\label{plusFourierAction}
\end{align}
where  
\begin{align}
d_0=2\log\left[\frac{\Gamma(\frac{1}{4})\Gamma(\frac{5}{4})}{\Gamma^2(\frac{3}{4})}\right]
\end{align}
and $\mathcal{G}$ is Catalan's constant:\begin{align}
\mathcal{G}=\sum_{k=0}^\infty \frac{(-)^k}{(2k-1)^2}=0.951597\dots.
\end{align}
Note that 
\begin{align}
0<d_0   \simeq 1.56638\dots.
\end{align}     

Eq. (\ref{plusFourierAction})  allows one to conclude that:
\begin{align} 
z^+_*(P)=\imath\frac{2\pi-P}{8(2\mathcal{G}-1)}+\dots
\end{align}
Putting this result back in the Fourier action gives:
\begin{align}
e^{-N(\varphi_+^{(0)}(z^+_*(P),P)+\imath z^+_*(P)P)}=e^{Nd_0-\frac{N(P-2\pi)^2}{16(2\mathcal{G}-1)}}       .
\end{align}
Such that the saddle point approximation for $e^{-NG_+^{(0)}}$ is given by:

\begin{align}
e^{-N\tilde{G}_+^{(0)}(P)}=\frac{1}{4\sqrt{\pi N(2\mathcal{G}-1)}     }e^{Nd_0-\frac{N(P-2\pi)^2}{16(2\mathcal{G}-1)}},
\end{align}

        Now we look at the part of the Fourier contour integral below the real axis.    
The action behaves around the points $\pm\frac{\imath}{2}$ as follows:
\begin{align}
\varphi_-^{(0)}(z)+\imath Pz\simeq\mp\frac{\imath \pi}{2}+\log\left(z\mp\frac{\imath}{2}\right) + \imath P z.
\end{align}
Such that the saddle point for large $P$ behave as:
\begin{align} 
z_*^{(1,2)}=\pm\frac{\imath}{2}+\frac{\imath}{P}
\end{align}
If $P$ is positive the steepest descent contour passes through $z_*^{(2)}$ and otherwise it passes through $z_*^{(1)}$, such that the action at the relevant saddle point, which we denote simply by $z^-_*$, is given by:
\begin{align}
\varphi_-^{(0)}(z_*)+\imath z^-_*P=\frac{|P|}{2}\,-\log(P)+\frac{\imath\pi}{2}\mp\frac{\imath \pi}{2}\dots
\end{align} 
Namely, the action is large and positive both at large positive and negative $P$.  A global minimum of the action occurs at $P_*^-=0$.  at which point $z_*^{(\pm)}(0)=0 $ (one can ascertain from (\ref{varphi0explicitCases}) that indeed $\varphi'^{(0)}_-(0)=0$). The fact that  the function $\tilde{G}$ obtains a minimum can be seen by expanding $\tilde G_-(P_*(z))=\varphi_-^{(0)}(z)+\imath z P(z)$ around $z=0$. One can check that this is a global minimum by further studying the function $\tilde{G}_-(P)$. The Fourier action then looks around $z=0$ and $P=0$ as follows:
\begin{align}
\varphi_-^{(0)}(z)+\imath zP=-d_0+\imath Pz+4(2\mathcal{G}-1)z^2+\dots,\end{align}
giving:
\begin{align}
z^+_*(P)=-\imath\frac{P}{8(2\mathcal{G}-1)}+\dots
\end{align}
One may now compute the saddle point approximation of $e^{-N\tilde{G}^{(0)}_-}$ just as before. In fact one may incorporate  the result already obtained for $e^{-N\tilde{G}^{(0)}_-}$ into the final result the encompasses   both  $\tilde{G}_+$ and $\tilde G_-$:

\begin{align}
e^{-N\tilde{G}^{(i)}(P)}=\frac{1}{4\sqrt{\pi N(2\mathcal{G}-1)}     }\sum_{\pm}  e^{Nd_i^\pm-\frac{N(P-P^\pm_*)^2}{16(2\mathcal{G}-1)}},
\end{align}
where 
\begin{align}
d^\pm_i =d_0-\Delta\varphi^{(i)}( 0 \pm\imath 0^+).  \end{align}
In this equation we further defined 
\begin{align}
\Delta\varphi^{(i)}\equiv\varphi^{(i)}-\varphi^{(0)} \label{Deltavarphidef}
\end{align}
and
\begin{align}
P_*^+ =2\pi, \quad P_*^-=0.
\end{align}
The Gaussian function behaves as a delta function for $P$ which varies on a scale of order $1$ or more (or, equivalently, $p$ varies on a scale of $N$ or more):
 
\begin{align}
e^{-N\tilde{G}^{(i)}(P)}\sim\frac{1}{N}\sum_{\pm} e^{Nd_i^\pm} \delta (P-P_*^\pm),
\end{align}
for small $P$, namely for $P\ll \frac{1}{\sqrt{N}}$ (or equivalently $p \ll \sqrt{N}$)the function $e^{-N\tilde{G}^{(i)}(P)}$ hardly varies and we may write:
\begin{align}
e^{-N\varphi^{(i)}_\pm(x)} \sim \delta_{\pm,-}     e^{Nd_i^-} \delta(x).\label{etovarphiisdelta}
\end{align}
This approximation is valid when $x$ varies on a scale larger than $\frac{1}{\sqrt{N}}$.

\section{Solution of the Integral Equation}
{\bf

}We define $e^{-N \tilde S(P,w)}$ as a function that  solves the following equation
\begin{align}
e^{-N \tilde S^{(i)}(P,w)}-\int_0^\infty NdQ e^{-N[\tilde{G}^{(i)}(P-Q)+\tilde S(Q ,w)+Q]}=\chi_{P>0} e^{-\imath N P w}, \label{TildeSEq}
\end{align}
an equation which is closely related to the Fourier transform of (\ref{Equation4R}).
Indeed, we shall later relate the solution of (\ref{TildeSEq}) to the solution of (\ref{Equation4R}). 

We solve Eq. (\ref{TildeSEq}) by positing the following leading order expressions for $e^{-N \tilde S^{(i)}(P,w)} $ for positive  $P$\  of order $1$:
\begin{align}
e^{-N \tilde S^{(i)}(P,w)} =\left\{\begin{array}{lr} e^{ -N[\imath P(w+\imath)-\varphi_-^{(i)}(w)]}  +\dots & 0<P<\Re [\Lambda(w)] \\ e^{ -N\imath Pw} +\dots & P>\Re [\Lambda(w)] \end{array} \right. ,\label{tildeSCases}
\end{align}
where here 
\begin{align}
\Lambda(w)=-\varphi^{(i)}_-(w).
\end{align}
To keep $\Lambda(w)$\ real we must take $w$ to be away from the real axis. The deviation from real $w$ is small and of order  $\frac{1}{N}$ , since $\varphi_-^{(0)}(x)$ is real for $x$ real. 

To ascertain that (\ref{tildeSCases}) provides a  solution to (\ref{Equation4R}) note that the $Q$ integral in Eq. (\ref{TildeSEq}) allows the integrand to pass through the saddle point. Indeed,  assuming Eq. (\ref{tildeSCases}), the integrand for $P<\Re \Lambda(w)$  is the exponent of the following expression, up to terms independent of $\tilde Q$ , which is defined as $\tilde{Q}=P-Q$:
\begin{align}
 -N[\tilde{G}^{(0)}(\tilde Q)-\imath \tilde Qw].
\end{align}
We have taken $i=0$, without loss of generality since corrections to that choice are of order $\frac{1}{N}$. The integral is to be performed over $\tilde {Q}<P$. For $w$ real the saddle point lies at imaginary $\tilde{Q}$, due to the symmetry $\tilde{G}^{(0)}(Q)=\overline{\tilde{G}^{(0)}}(-Q)$, which is due in turn to the fact that $\varphi_-^{(0)}$ is real on the real axis. This entails that $\tilde{G}^{(0)}$ is real on the imaginary axis, where the saddle point is to be found. The integral may now be performed by noting that if the integral is not restricted to positive $\tilde{Q}<P$ we have:
\begin{align}
N\int d{\tilde Q}e^{-N[\tilde{G}^{(i)}(\tilde Q)-\imath \tilde Qw]}=e^{-\varphi^{(i)}(w)}
\end{align} 
and the restriction to $\tilde{Q}<P$ does not matter in the saddle point approximation, since the saddle point on the imaginary axis is reached by deforming the path of integration of $\tilde{Q}$ for positive $P$.

 Now we shall consider the case left out in Eq. (\ref{tildeSCases}), namely $P<0$ and of order $1$. We have:
\begin{align}
e^{-N \tilde S^{(i)}(P,w)}=\int_0^\infty NdQ e^{-N[\tilde{G}^{(i)}(P-Q)+\tilde S(Q ,w)+Q]},\quad P<0.\label{negativeP}
\end{align}
Note that the right hand side of this equation may be computed once $\tilde{S}(P,w)$ is known for positive $P$, e.g., by using Eq. (\ref{tildeSCases}) for $Q$\  of order 1. Nevertheless, the integral in (\ref{negativeP}) does not pass through the saddle point, and thus always give a subdominant contribution. The outcome is that it will not appear in the calculation below.

Given $e^{-N \tilde S^{(i)}(P,w)}$ for arbitrary $w$,  we may easily construct a solution of (\ref{Equation4R}).
First, given $\tilde S,$\ we construct $\mathcal{M}$ as follows:
\begin{align}
\mathcal{M}^{(i)}(z,w) = \frac{1}{z-w}+\int_{-\infty}^\infty NdP\left(e^{-N\tilde{S}^{(i)}(P,w)}-\chi_{P>0}e^{ -N\imath Pw}\right),
\end{align}
or in Fourier space:
\begin{align}
\mathcal{M}^{(i)}(P,w)=-\chi_{P<0}e^{-N\imath P w}+e^{-N\tilde{S}^{(i)}(P,w)},
\end{align}
where it is assumed the $w$ is on the real axis. 

The function $\mathcal{M}$ solves the equation:
\begin{align}\mathcal{M}^{(i)}(z,w) - \oint_{\mathcal{C}} e^{-N\varphi^{(i)}(z')}   \frac{\mathcal{M}^{(i)}(z'+\imath,w)dz'}{(z-z')2\pi\imath} = \frac{1}{z-w}.\label{CrulyMExpression}
\end{align}
Thus  $\mathcal{R}$ is given by 
\begin{align}
\mathcal{R}^{(i)}(z,w) =\mathcal{M}^{(i)}(z,w) - \gamma ^{(i)}(w)\mathcal{M}^{(i)}\left(\ z,-g^{-1}-  \frac{\imath}{2}\right),\label{RtwoMs} 
\end{align}
where $\gamma^{(KK)}=0,$ while $\gamma^{(NK)}$ is given by:
\begin{align}
\gamma ^{(NK)}(w)=\frac{\mathcal{M}^{(NK)}(-g^{-1}+\frac{\imath}{2},w) }{\mathcal{M}^{(NK)}\left(\ -g^{-1}+\frac{\imath}{2},-g^{-1}-  \frac{\imath}{2}\right)}
\end{align}
such that  Eq. (\ref{gammacondition}) is satisfied.

Now we consider the case where  $|P|\ll1$. In this case we may treat $e^{-N\varphi_-(z)}$ { as a delta function concentrated around $z=0$} (see Eq. (\ref{etovarphiisdelta})):
\begin{align}
e^{-N\varphi_-^{(i)}(z)} =e^{N d^-_i}\delta (z)
\end{align}
and thus we have the following solution of Eq.  (\ref{CrulyMExpression}) for $\mathcal{M}$:
\begin{align}
\mathcal{M}^{(i)}(z,w)=\frac{1}{z-w}+\imath\frac{1-2\pi e^{-Nd^-_i}}{z(w-\imath)}.\label{lowPm}
\end{align}

\section{Computation of the Nonequilibrium Sources}

In \cite{Bettelheim:Kondo} we wrote a form for the variation of the  logarithm of the determinant in Eq. (\ref{OverlapasDet}), but we shall use here a different form of this variation, which we derive below in the first subsection, while in the next subsection we shall used the derived expression in order to compute the non-equilibrium sources. The nontrivial part of this calculation is the logarithm of the determinant in  Eq. (\ref{OverlapasDet}), to which we shall devoted our attention.

First, let us note that we are interested in computing:
\begin{align}
\delta_{\sigma^{(\bm u)}} \log\frac{|\<\bm u|\bm v\>|}{\sqrt{|\<\bm u|\bm u\>||\<\bm v|\bm v\>|}},
\end{align}
where $\delta_{\sigma^{(\bm u)}}$ relates to the variation of the  rapidities $v_i$.  This variation is given, through Eq. (\ref{OverlapasDet}), by the following:
\begin{align}
\delta_{\sigma^{(\bm u)}} \log\frac{|\<\bm u|\bm v\>|}{\sqrt{|\<\bm u|\bm u\>||\<\bm v|\bm v\>|}}=\Re  \lim_{\bm u'\to\bm u}\tr\left[(\delta_{\sigma^{(\bm u)}}B^{(\bm u\cup\bm v)}  )B^{(\bm u\cup\bm v)-1}-(\delta_{\sigma^{(\bm u)}}B^{(\bm u'\cup\bm u)}  )B^{(\bm u'\cup\bm u)-1}\right].
\end{align}  
For brevity, using the notation of Eq. (\ref{brevity}), we shall write this as:
\begin{align}
\delta_{\sigma^{(\bm u)}} \log\frac{|\<\bm u|\bm v\>|}{\sqrt{|\<\bm u|\bm u\>||\<\bm v|\bm v\>|}}=\Re\ \tr\left[ (\delta_{\sigma}B^{(i)}  )B^{(i)-1}  \right]^{i=NK}_{i=KK}.\label{wnanttocompute}
\end{align}
In fact, the $\dots^{i=NK}_{i=KK}$ will always denote subtracting the result of $\dots$ when $KK$ is substituted for $i$  from the result when $NK$ is substituted from $i$, and the variation $\delta_\sigma$ will always suggest a variation of $\bm u$, where for $i=KK$ a limiting procedure is implied as above. Namely, $\bm u$ and $\bm u'$ are first to be employed, then the variation is to be taken and finally $\bm u'$ is to be let to tend to $\bm u$.

\subsection{Variation of the Logarithm of the Determinant}

To find trace appearing in (\ref{wnanttocompute}), we first note that we are interested in the trace of the following object:

\begin{align}
&(\delta B^{(l)}B^{(l)-1} )_{ik}=\sum_{j} \delta_\sigma \left( \frac{Q^{(l)}_{\bm c}(c_i+\imath)}{Q_{\bm z}(c_i+\imath)}\frac{Q_{\bm z}(c_i)}{Q'^{(l)}_{\bm c}(c_i)} \frac{1}{c_i-c_j+\imath } \right) B^{(l)-1}_{jk} 
\end{align}
(to avoid confusion we remark that $B^{(l)-1}$ is the inverse of $B^{(l)}$, where $B^{(l)}$ is given in Eq.(\ref{brevity})).  
More explicitly, making use of Eqs. (\ref{Bexpression}) and (\ref{Rdefinition}) we may write:
\begin{align}
&(\delta B^{(l)}B^{(l)-1} )_{ik}= \mathcal{R}^{(l)}(c_i+\imath,c_k)\delta_\sigma\frac{Q^{(l)}_{\bm c}(c_i+\imath)}{Q_{\bm z}(c_i+\imath)}\frac{Q_{\bm z}(c_i)}{Q'^{(l)}_{\bm c}(c_i)} +\label{deltaBInversBdiscrete}
\\&+\nonumber\frac{Q^{(l)}_{\bm c}(c_i+\imath)}{Q_{\bm z}(c_i+\imath)}\frac{Q_{\bm z}(c_i)}{Q'^{(l)}_{\bm c}(c_i)} \left(\delta c_i \mathcal{R}'^{(l)}(c_i+\imath,c_k)+\oint \frac{f_{\delta \sigma}(s) \mathcal{R}^{(l)}(s,c_k)}{(c_i+\imath-s)^2} \frac{ds}{2\pi\imath} \right),
\end{align} 
where $\mathcal{R}'(z,w)\equiv \partial_z \mathcal{R}(z,w)$.

We compare (\ref{deltaBInversBdiscrete}) with the following double contour integral
where the $w$ contour surrounds the $z$ contour, which surrounds the real axis, we drop the superscript $(l)$ for brevity for this calculation:

\begin{align}
&\oiint \frac{dwdz}{(2 \pi \imath)^2 (w-z)} \mathcal{R}(z+\imath,w)\delta_\sigma  e^{-N\varphi(z)}=\\&=\left.\oint \frac{dw}{2\pi \imath} \delta_\sigma\oint \frac{dz}{2\pi\imath(z-w)} \mathcal{R}^{(\sigma')}(z+\imath,w) e^{-N\varphi_\sigma(z)} \right|_{\sigma'=\sigma}\nonumber
 \end{align}
 Taking the variation with respect to $\sigma$ explicitly, one obtains:
\begin{align} 
&\oiint \frac{dwdz}{(2 \pi \imath)^2 (w-z)} \mathcal{R}(z+\imath,w)\delta_\sigma  e^{-N\varphi(z)}=\\&=\sum_{i} \oint \frac{dw}{2\pi \imath } \frac{\mathcal{R}(c_i+\imath,w)}{w-c_i}\delta_\sigma\frac{Q^{(i)}_{\bm c}(c_i+\imath)}{Q_{\bm z}(c_i+\imath)}\frac{Q_{\bm z}(c_i)}{Q'^{(i)}_{\bm c}(c_i)} +\frac{Q^{(i)}_{\bm c}(c_i+\imath)}{Q_{\bm z}(c_i+\imath)}\frac{Q_{\bm z}(c_i)}{Q'^{(i)}_{\bm c}(c_i)}\delta c_i \partial_{c_i}\frac{\mathcal{R}(c_i+\imath,w)}{w-c_i}\nonumber,\end{align}
where the sum is only over those $i$'s for which $c_i$ is on the real axis. Namely, excluding $c_{N+1}$ if $l=NK$. 
The $w$ integral may now be performed to obtain:
\begin{align}
&\oiint \frac{dwdz}{(2 \pi \imath)^2 (w-z)} \mathcal{R}(z+\imath,w)\delta_\sigma  e^{-N\varphi(z)}\label{doubleintegralresult}=\\&=\sum_{i}  \mathcal{R}(c_i+\imath,c_i)\delta_\sigma\frac{Q^{(i)}_{\bm c}(z+\imath)}{Q_{\bm z}(z+\imath)}\frac{Q_{\bm z}(z)}{Q'^{(i)}_{\bm c}(z)} +\frac{Q^{(i)}_{\bm c}(c_i+\imath)}{Q_{\bm z}(c_i+\imath)}\frac{Q_{\bm z}(c_i)}{Q'^{(i)}_{\bm c}(c_i)}\delta c_i \partial_{c_i}\mathcal{R}(c_i+\imath,c_i)\nonumber
 \end{align}
where the integral over $w$ surrounds the integral over $z$ and $\mathcal{R}'$ denotes a derivative of $\mathcal{R}$ with respect to the first argument.

Let us now compute the variation related to the diagonal $N+1$ element in the case where $B^{(i)}$ in question is $B^{(NK)}$. The rapidity $c_{N+1}$ in this case is $c_{N+1}=-g^{-1}-\frac{\imath}{2}+\delta$. Namely, this rapidity does not lie on the real axis and,and is thus  missed in the treatment above. A similar calculation to the one just performed, keeping in mind that the $N+1$ rapidity is kept fixed ( $\delta c_{N+1}=0$) gives:
\begin{align} 
&(\delta _\sigma B^{(NK)}B^{{(NK)}-1} )_{N+1,N+1}=\underset{z\to c_{N+1}}{\Res}e^{-N\varphi(z)}\left(-N \delta_\sigma\varphi(c_{N+1})   \mathcal{R}^{(NK)}(c_{N+1}+\imath,c_{N+1})  +\nonumber\right. \\&+\label{DiagonalContribution} \left. \oint\frac{f_{\delta \sigma}(s) \mathcal{R}^{(NK)}(s,c_{N+1})}{(c_{N+1}+\imath-s)^2} \frac{ds}{2\pi\imath}\right)
\end{align}
Here it is implied that the limit $\delta =c_{N+1}-z_1+\imath\to0$ has not been taken. The integral on the second line is order $1$ and can be dismissed, as we shall see that the leading order correction is of order $N$. To see that the integral on the second line is indeed only of order $1$ consider that a factor of order $N$ can only come from large momenta, of order $N$ ($p\sim N$ or $P\sim 1$)   of the numerator in the integrand.   $f_{\delta \sigma}$ is assumed to have only low momentum content, then the large momenta can come only from  $\mathcal{R}. $  However, such large momentum cannot be offset against any other term in the integral and thus there is no such contribution. 

We now turn to the first term in (\ref{DiagonalContribution}). Due to Eq. (\ref{gammaessenece}) it may be recognized as:
\begin{align}
N\delta_\sigma\varphi(c_{N+1}) \gamma^{(NK)}\left(-g^{-1}-\frac{\imath}{2}\right)
\end{align} 
Since $\gamma^{(NK)}\left(-g^{-1}-\frac{\imath}{2}\right)=1$ we conclude that to leading order :
\begin{align} 
&(\delta _\sigma B^{(NK)}B^{{(NK)}-1} )_{N+1,N+1}=N\delta_\sigma\varphi(c_{N+1})\label{diagnoalNp1variation} 
 \end{align}

Comparing (\ref{deltaBInversBdiscrete},\ref{doubleintegralresult})  and (\ref{diagnoalNp1variation}) one obtains
\begin{align}
&\left.\tr \delta_\sigma B^{(i)}B^{(i)-1} \right|^{i=NK}_{i=KK}=\oiint \frac{dwdz}{(2 \pi \imath)^2 (w-z)}\mathcal{R}^{(i)}(z+\imath,w)\delta_\sigma  e^{-N\varphi(z)}\nonumber+\\&-\oiint \frac{dwdze^{-N\varphi^{(i)}(z)}}{(2 \pi \imath)^2 (w-z)}\left[f_{\delta \sigma}(z)\partial_w\mathcal{R}^{(i)}(z+\imath,w)-\oint\frac{ds}{2\pi\imath} \frac{f_{\delta \sigma}(s) \mathcal{R}^{(i)}(s,w)}{(z+\imath-s)^2}\right]^{i=NK}_{i=KK} \nonumber+\\
&+N\delta _\sigma\varphi^{(0)}\left(-\frac{\imath}{2}-g^{-1}\right),\label{conventierform}
\end{align}
 to leading order, which as we shall confirm later is $O(N)$. The $w$ integral surrounds the $z$ integral which surrounds, in turn, the $s$\ integral.

\subsection{Computation}   
Having the variation of the determinant in terms of $\mathcal{R}$ and a solution for $\mathcal{R}$, we proceed to compute the variation of the determinant. To this purpose, we first define 
\begin{align}
m^{(i)}(w,P)=e^{-N\left[\tilde{S}(P,w) -\imath P(w+\imath)+\varphi_-^{(i)}(w)\right]}.
\end{align}
From (\ref{tildeSCases}) we have that 
\begin{align}
m^{(i)}(w,P) = \left\{\begin{array}{lr}1 & 0<\varphi^{(i)}(w)-P\sim\ 1 \\0  &0<P-\varphi^{(i)}(w)\sim 1\end{array} \right. \label{mbehavior}.
\end{align}
and 
\begin{align}
m^{(i)}(w,P) =m^{(0)}(w,P-\Delta\varphi^{(i)}(w)), 
\end{align}
for $P,P-\Delta\varphi^{(i)}(w)>0$ and $\Delta\varphi$ is defined in (\ref{Deltavarphidef}).

From which one may conclude
\begin{align}
\int _0^\infty m^{(i)}(w,P)dP=\varphi^{(i)}(w)+c,\label{mdP} 
\end{align}
where $c\ll1$, according to (\ref{mbehavior}). We can also compute the following integral which will appear later:
\begin{align}
&\int _0^\infty m^{(i)}(w,P) P dP=\int _{\max(0,\Delta\varphi^{(i)})}^\infty  m^{(0)}(w,P-\Delta\varphi^{(i)}(w))PdP=\label{mPdP}
\\&=\int _{0}^\infty  m^{(0)}(w,P)(P+\Delta\varphi^{(i)}(w))dP+o(1/N)=\nonumber\\&=\varphi_-^{(0)}(w)\Delta  \varphi_-^{(i)}(w)+\tilde c+o(1/N),\nonumber 
\end{align}
where $\tilde c$ is independent of $i$ and thus an unimportant constant, as will become apparent  later.

First concentrate on the  contribution to (\ref{conventierform}) obtained by substituting $\mathcal{M}^{(i)}(z,w)$ for $\mathcal{R}^{(i)}(z,w)$, reserving for later the substitution of $\gamma^{(i)}(w)\mathcal{M}^{(i)}(z,-g^{-1}-\imath) $ , which from Eq. (\ref{RtwoMs}) constitutes the second part of $\mathcal{R}^{(i)}(z,w)$. We have for the contribution currently under consideration:\begin{align}
& N\oint\frac{dw}{(2\pi\imath)^2}\int m^{(i)}(w,P) dP \int dz  \frac{e^{-N\varphi_-^{(i)}(z)}}{w-z}\times \\&\times\left. \left[ N   \frac{\delta\varphi_-^{(i)}(z) }{\delta \sigma}-\partial_z f_{\delta \sigma}(z+\imath)-f_{\delta \sigma}(z) 
\partial_w\right] \nonumber e^{N [\imath P(z-w)+\varphi_-^{(i)}(w)]}\right|^{i=NK}_{i=KK}=\\&=\left. N^2 \int\frac{dz}{2\pi\imath}\int  m^{(i)}(z,P)dP  \left[    \frac{\delta\varphi_-^{(i)}(z)}{\delta  \sigma}-f_{\delta \sigma}(z)\varphi'^{(i)}_-(z)+ 
\imath P(f_{\delta \sigma}(z)-f_{\delta \sigma}(z+\imath))\right] \nonumber \right|^{i=NK}_{i=KK}.
\end{align}
Here we have taken $z$ integration to be solely under the real axis. This is because    the momentum content of $e^{-N\varphi_+^{(i)}(z)}$ lies around $P=2\pi$, where $\mathcal{M}^{(i)}$ already decays, thus  giving a negligible contribution. The part of the integration contour of  $z$ which is  below the real axis where $e^{-N\varphi_+^{(i)}(z)}$contributes non-trivially to the double integral as we shall see. 

We can further use Eq. (\ref{mPdP}) and (\ref{mdP}) to write the following expression for the contribution under current consideration: 
\begin{align}
 \label{almostfinal}N^2\int \left[\varphi_-^{(i)}(x)  \left[    \delta_\sigma \varphi_-^{(i)}(x)-f_{\delta \sigma}(x)\varphi'^{(i)}_-(x)\right] +\varphi_-^{(0)}(x)  \varphi_-^{(i)}(x)(f_{\delta \sigma}(x)-f_{\delta \sigma}(x+\imath))\right]^{i=NK}_{i=KK}dx.
\end{align}
We have kept only terms of order $N$ (the integral is of order $1/N$ after taking the difference between $i=NK$ and $i=KK$). It is at this point at which one observes that terms related to $\tilde c$ of Eq. (\ref{mPdP}) drop out to leading order in $N$.

We should also check for  the low $P$ behavior of the same contribution, but when inserting (\ref{lowPm}) for $\mathcal{M}$ one obtains an expression which is only of order $1$ and thus negligible as compared to (\ref{almostfinal}):    
\begin{align}
&  N\int \frac{dz}{2\pi\imath}\int\frac{dw}{2\pi\imath}  \frac{e^{Nd^-_i}}{z-w}{\delta{_\sigma }} d_i\delta(z) \left[ \frac{1}{z+\imath-w}+\imath\frac{1-2\pi e^{-Nd^-_i}}{(z+\imath)(w-\imath)}
\right]^{i=NK}_{i=KK}=\\&=N\left. {\delta{_\sigma }} d_i\right|^{i=NK}_{i=KK}=O(1). \nonumber
\end{align}

We now turn to $\gamma^{(i)}(w)\mathcal{M}^{(i)}(z,-g^{-1}-\imath) $ . The integral thus obtained has the following form in the low $P$ limit:
\begin{align}
&  \int \frac{dz}{2\pi\imath}\int\frac{dw}{2\pi\imath}  \frac{e^{Nd^-_i} \delta(z) }{w-z} \delta{_\sigma } d_i^-\frac{\mathcal{M}^{(NK)}(-g^{-1}+\frac{\imath}{2},w)\mathcal{M}^{(i)}\left(\ z+\imath,-g^{-1}-  \frac{\imath}{2}\right) }{\mathcal{M}^{(NK)}\left(\ -g^{-1}+\frac{\imath}{2},-g^{-1}-  \frac{\imath}{2}\right)} \nonumber =\\
 &=N \int \frac{dz}{(2\pi\imath)^2}  e^{Nd^-_i} {\delta{_\sigma }} d_i\frac{\mathcal{M}^{(NK)}(-g^{-1}+\frac{\imath}{2},0)\mathcal{M}^{(i)}\left(\ \imath  ,-g^{-1}-  \frac{\imath}{2}\right) }{\mathcal{M}^{(NK)}\left(\ -g^{-1}+\frac{\imath}{2},-g^{-1}-  \frac{\imath}{2}\right)} \nonumber=\\
 &=O(e^{-Nd^-_i}).
\end{align}
The large $P$\ limit similarly has an exponentially small contribution and is also discarded.

Our main result then is:
\begin{align}
&\delta_{\sigma^{(\bm u)}} \log\frac{|\<\bm u|\bm v\>|}{\sqrt{|\<\bm u|\bm u\>||\<\bm v|\bm v\>|}}=N\Re\left[\delta _\sigma\varphi^{(0)}\left(-\frac{\imath}{2}-g^{-1}\right)\right.+\label{FinalResult}
\\&+\left.N\int dx \left(f_{\delta \sigma}(x)-f_{\delta \sigma}(x+\imath)+\delta_\sigma-f_{\delta \sigma}(x)\partial_x\right) \varphi_-^{(0)}(x)\varphi^{(i)}_-(x)|^{i=NK}_{i=KK}\right]\nonumber,
\end{align}
where we have added to (\ref{almostfinal})
the last term in (\ref{conventierform}). 

\section{Conclusion}
The purpose of this paper is to obtain the behavior after a quench from an initial normal metal state to the Kondo state at long times. As explained in the introduction, to find this behavior,  Eq. (\ref{QActionEq}) must be solved. An expression for the left hand side of that equation is familiar from equilibrium physics. The right hand side, namely the functional derivative $ \frac{\delta \mathcal{A}}{\delta \bm \sigma}$,  must be calculated, whereupon it serves as a source term in the equilibrium equation. The functional derivative, $ \frac{\delta \mathcal{A}}{\delta \bm \sigma},$ was computed here to be given by Eq. (\ref{FinalResult}), the main result of this paper, plus a source term associated with the charge sector, which is much simpler to compute and may be obtained by looking up  Ref. \cite{Bettelheim:Kondo}.  To make use of the result of this equation,   Eq. (\ref{FinalResult}), one must first compute the postulate a change of density $\delta \sigma$ associated with an excitation over the Kondo ground state. After postulating  $\delta\sigma$ the motion of the individual rapidities, which  is described by $f_{\delta\sigma}$, must be calculated. This is facilitated by  the relation between  $\delta\sigma $ and $f_{\delta\sigma}$  given in Eq. (\ref{fdeltasigmarelation}) . With these at hand, all the elements of the right hand side (\ref{FinalResult}) are given, and one may proceed to solve Eq. (\ref{QActionEq}).

We leave the solution of Eq.  (\ref{QActionEq}), for different initial states and for different values of $\gamma$ to future work, restricting the purpose of this paper to obtaining the more general equations from which such solutions may be obtained. 

Our approach may be compared to other approaches. For example one may in particular study the similarities to the developments presented in Refrs. \cite{Saleur:Lukyanov:Overlaps:Kondo,Lesage:Saleur:Boundary:Impurity:Problems,Saleur:Corssover:In:Impurity,Lesage:Saleur:Boundary:Conditions:Changing:Non:Conformal:Theories}, which are based on the axiomatic approach. It seems that the approach presented here may be complimentary to approach presented in the latter references. A thorough comparison, which will also serve to validate the current approach is left to future work. Nevertheless, we note that the current approach has the potential of being generalizable to other cases in which the initial and final states may not be close to the ground state. In such a case, the solution of the integral equation central to our approach, Eq. (\ref{Equation4R}), has to be of course  reconsidered for the particular case at hand. The current paper then serves as a starting point for such investigations.

\section{Acknowledgement  }
I acknowledge
the Israeli Science Foundation, which supported this research through grant 1466/15.


\end{document}